\documentclass{icrc2009}

\usepackage{graphicx}   
\usepackage[caption=false]{caption}    
\usepackage[font=footnotesize]{subfig} 
\usepackage{fixltx2e}
\usepackage{url}

\newcommand{\shorttitle}[1]%
{\markboth{Proceedings of the 31\MakeLowercase{$^{st}$} ICRC, {\L}\'{o}d\'{z} 2009}{#1} }
\newcommand{\etal}{\MakeLowercase{\textit{et al. }}} 

\usepackage{epstopdf}
\usepackage{amssymb,amsmath}

\begin{document}
\title{AMANDA 7-Year Multipole Analysis}

\author{\IEEEauthorblockN{Anne Schukraft\IEEEauthorrefmark{1},
			  Jan-Patrick H{\"u}l\ss\IEEEauthorrefmark{1},
			  Christopher Wiebusch\IEEEauthorrefmark{1} for the IceCube Collaboration\IEEEauthorrefmark{2}}
                            \\
\IEEEauthorblockA{\IEEEauthorrefmark{1}III. Physikalisches Institut, RWTH Aachen University, Germany}
\IEEEauthorblockA{\IEEEauthorrefmark{2}see http://www.icecube.wisc.edu/collaboration/authorlists/2009/4.html}}

\shorttitle{Schukraft \etal AMANDA Multipole Analysis}
\maketitle

\begin{abstract}
The multipole analysis investigates the arrival directions of registered neutrino events in AMANDA-II by a spherical harmonics expansion. The expansion of the expected atmospheric neutrino distribution returns a characteristic set of expansion coefficients. This characteristic spectrum of expansion coefficients can be compared with the expansion coefficients of the experimental data. As atmospheric neutrinos are the dominant background of the search for extraterrestrial neutrinos, the agreement of experimental data and the atmospheric prediction can give evidence for physical neutrino sources or systematic uncertainties of the detector. Astrophysical neutrino signals were simulated and it was shown that they influence the expansion coefficients in a characteristic way. Those simulations are used to analyze deviations between experimental data and Monte Carlo simulations with regard to potential physical reasons. The analysis method was applied on the AMANDA-II neutrino sample measured between 2000 and 2006 and results are presented.
  \end{abstract}

 
\section{Introduction}
The AMANDA-II neutrino detector located at South Pole was constructed to search for astrophysical neutrinos. These neutrinos could originate from many different Galactic and extragalactic candidate source types such as Active Galactic Nuclei (AGN), supernova remnants and microquasars.
The detection of neutrinos is based on the observation of Cherenkov light emitted by secondary muons produced in charged current neutrino interactions. This light is observed by photomultipliers deployed in the Antarctic ice. Their signals are used to reconstruct the direction and the energy of the primary neutrino.

AMANDA-II took data between 2000 and 2006. The background of atmospheric muons is reduced by selecting only upward-going tracks in the detector, as only neutrinos are able to enter the detector from below. This restricts the field of view to the northern hemisphere.

The data is filtered and processed to reject misreconstructed downward-going muon tracks \cite{JIMBRAUN}. The final data sample contains 6144 neutrino induced events between a declination of $0^{\circ}$ and $+90^{\circ}$ with a purity of $> 95\%$ away from the horizon.\\

\section{Analysis Principle}

The idea of this analysis is to search for deviations of the measured AMANDA-II neutrino sky map from the expected event distribution for atmospheric neutrinos, which constitute the main part of the data sample \cite{MUL4YEAR}. A method to study such anisotropies is a multipole analysis, which was also used to quantify the Cosmic Microwave Background fluctuations.
The analysis is based on the decomposition of an event distribution $f(\theta, \phi) = \sum_{i=1}^{N_\mathrm{events}} \delta(\cos\theta_i - \cos\theta) \delta(\phi_i - \phi)$ into spherical harmonics $Y_l^m(\theta, \phi)$, where $\theta$ and $\phi$ are the zenith and azimuth of the spherical analysis coordinate system. The expansion coefficients are
\begin{equation}
a_l^m = \int_0^{2\pi} d\phi \int_{-1}^1 d\cos\theta \ f(\theta, \phi) Y_l^{m*} (\theta, \phi).
\end{equation}
They provide information about the angular structure of the event distribution $f(\theta, \phi)$. The index $l$ corresponds to the scale of the angular structure $\delta \approx \frac{180^{\circ}}{l}$ while $m$ gives the orientation on the sphere. The expansion coefficients with $m=0$ depend only on the structure in the zenith direction of the analysis coordinate system. Averaging over the orientation dependent $a_l^m$ yields the multipole moments
\begin{equation}
C_l = \frac{1}{2l+1} \sum_{m=-l}^{+l} |a_l^m|^2.
\end{equation}
They form an angular power spectrum characteristic for different input neutrino event distributions.

The initial point of this analysis is the angular power spectrum of only atmospheric neutrino events. Therefore, neutrino sky maps containing 6144 atmospheric neutrino events according to the Bartol atmospheric neutrino flux model \cite{BARTOL} are simulated and numerically decomposed with the software package GLESP \cite{GLESP}. Statistical fluctuations are considered by averaging over 1000 random sky maps, resulting in a mean $\langle C_l \rangle$ and a statistical spread $\sigma_{C_l}$ of each multipole moment.

The same procedure is applied to simulated sky maps containing atmospheric and different amounts of signal neutrinos with a total event number of likewise 6144 events. The influence of the signal neutrinos on the angular power spectrum is studied in terms of the pulls
\begin{equation}
d_l = \frac{\langle C_l \rangle - \langle C_{l,\mathrm{atms}} \rangle}{\sigma_{C_{l,\mathrm{atms}}}}.
\label{pulls}
\end{equation}

 \begin{figure*}[!t]
   \centerline{\subfloat[]{\includegraphics[width=3.3in]{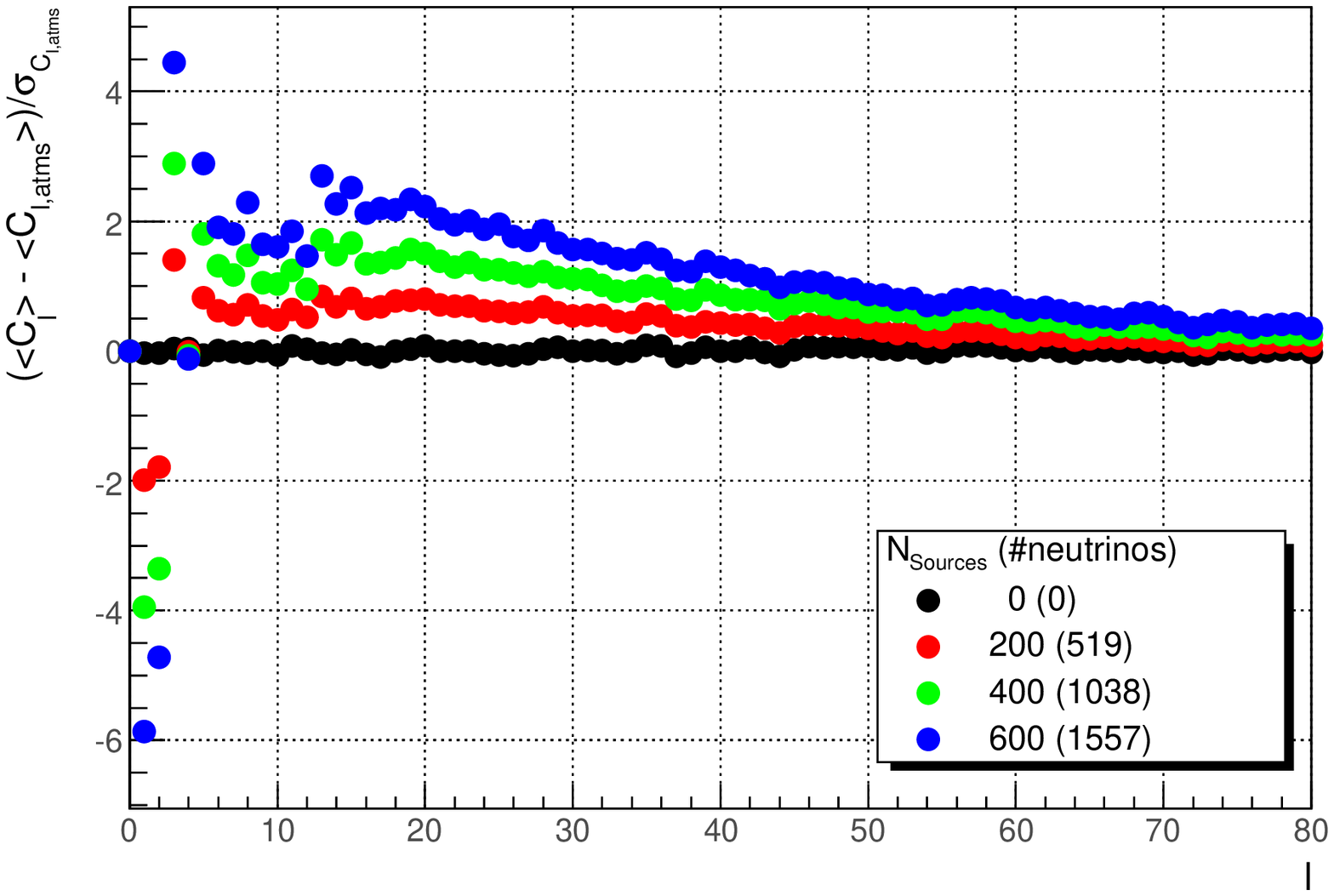} \label{sub_fig1}}
              \hfil
              \subfloat[]{\includegraphics[width=3.3in]{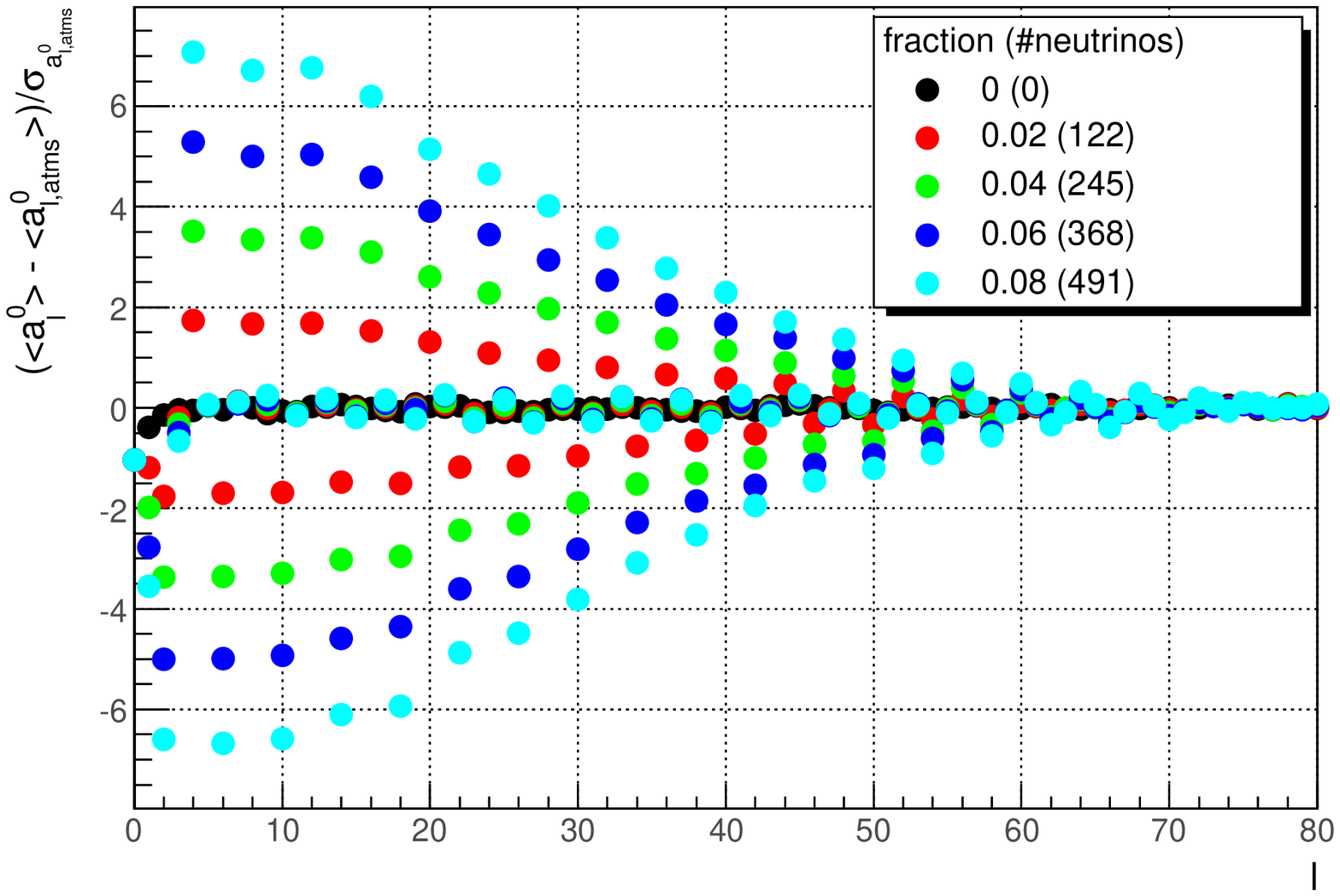} \label{sub_fig2}}
             }
   \caption{(a): Pull plot for the multipole moments $C_l$ of the isotropic point source model. Sources are simulated with a mean source strength $\mu = 5$ and an $E_{\nu}^{-2}$ energy spectrum. The number of sources $N_\mathrm{sources}$ on the full sphere is varied. The corresponding number of signal neutrinos on the northern hemisphere is given in brackets. The errors bars are hidden by the marker symbols. (b): Pull plot for the expansion coefficients $a_l^0$ of the cosmic ray interaction model with the Galactic plane in Galactic coordinates. The fraction of neutrinos in the sky map originating from the Galactic plane is varied. The corresponding number of signal neutrinos is given in brackets. The errors bars are hidden by the marker symbols.}
   \label{double_fig}
 \end{figure*}

\section{Signal simulation}
The different models for candidate neutrino sources investigated in this analysis are:
  \begin{enumerate}
   \item Isotropically distributed point sources
   \item A diffuse flux from FR-II galaxies and blazars \cite{BECKER}
   \item AGN registered in the V{\'e}ron-Cetty and V{\'e}ron (VCV) catalog \cite{VERON}
   \item Galactic point sources such as supernova remnants or microquasars
   \item Cosmic rays interacting in the Galactic plane.
  \end{enumerate}
All simulated pointlike neutrino sources are characterized by a Poissonian distributed source strength with mean $\mu$ and an energy spectrum $E_{\nu}^{-\gamma}$. The relative angular detector acceptance depends on the neutrino energy and therefore on the spectral index of the simulated neutrino source. Signal neutrinos are simulated according to this acceptance considering systematic fluctuations. The total number of signal neutrinos in a sky map of the northern hemisphere with $N_\mathrm{sources}$ simulated sources on the full-sky is therefore given by $\sim 0.5 \cdot \mu \cdot N_\mathrm{sources}$. Additionally the angular resolution is taken into account. It dominates over the uncertainty between the neutrino and muon direction.

The spectral index of pointlike sources is varied between $1.5 \le \gamma \le 2.3$. As the spectral index of atmospheric neutrinos is close to $3.7$, signal and backround neutrinos underlie different angular detector acceptances. Thus, additionally to the clustering of events around the source directions also the shape of the total angular event distribution is used to identify a signature of signal neutrinos in the angular power spectrum \cite{DIPLOMATHESIS}.

Neutrinos from our Galaxy disturb the atmospheric event distribution by their bunching within the Galactic plane modeled by a Gaussian band along the Galactic equator. Neutrinos produced in cosmic ray interactions with the interstellar medium of our Galaxy are assumed to follow the $E^{-2.7}$ primary energy spectrum.

A further topic (model 6) that can be studied with a multipole analysis are neutrino oscillations. The survival probability of atmospheric muon neutrinos depends on the neutrino energy and the traveling length of the neutrino as well as the mixing angle $\theta_{23}$ and the squared mass difference $\Delta m^2_{23}$. The traveling length can be expressed by the Earth's radius and the zenith angle of the neutrino direction \cite{DIPLOMATHESIS}. Thus, the neutrino oscillations disturb the angular event distribution of atmospheric neutrinos. With the assumption of $\sin^2 \left( 2\theta_{23} \right) \approx 1$ the squared mass difference remains for investigation. Due to the relatively high energy threshold of $50$ GeV the effect is small. \\

\section{Evaluation of the power spectra}

The deviations from a pure atmospheric angular power spectrum caused by signal neutrinos are studied by the pulls. These pulls are exemplarily shown in Fig. \ref{sub_fig1} for the model of isotropic point sources. The behaviour of the pulls is characteristic for each signal model. Different multipole moments carry different sensitivity to the neutrino signal. The absolute value of the pull increases linearly with the amount of signal neutrinos in the sky maps. Each pull has a predefined sign. 

The deviation of a particular sky map with multipole moments $C_l$ from the pure atmospheric expectation $\langle C_{l,\mathrm{atms}} \rangle$ is quantified by a significance indicator $D^2$ defined as
\begin{equation}
D^2 = \frac{1}{l_\mathrm{max}} \sum_{l=1}^{l_\mathrm{max}} \textrm{sgn}_l \cdot w_l \cdot \left( \frac{C_l - \langle C_{l,\mathrm{atms}} \rangle}{\sigma_{C_{l,\mathrm{atms}}}} \right)^2,
\label{significanceindicator}
\end{equation}
where $l_\mathrm{max}$ determines the considered multipole moments. The term in brackets is the pull between the particular sky map and the mean of the atmospheric expectation as defined in Eq. \ref{pulls}. The factors
\begin{equation}
w_l = \frac{\langle C_l \rangle - \langle C_{l,\mathrm{atms}} \rangle}{\sqrt{\sigma_{C_l}^2 + \sigma_{C_{l,\mathrm{atms}}}^2}}
\end{equation}
are defined to weight the pulls according to their expected sensitivity to the signal. For each neutrino signal model one dedicated set of weights ${w_l}$ is determined. Due to the linear increase of the pulls with the signal strength the strength chosen to calculate $w_l$ is arbitrary. 

The weight factors $w_l$ carry the expected sign of the pulls. sgn$_l$ is the sign of the measured pull. Thus, the $D^2$ calculated for the particular sky map is increased if the observed deviation has the direction expected for the signal model and reduced otherwise.

Due to the weighting of the pulls, the sensitivity becomes stable for high $l_\mathrm{max}$. A choice of $l_\mathrm{max} = 100$ is sufficient to provide best sensitivity to all investigated signal models. 

The $D^2$ of a sky map is interpreted physically by the use of confidence belts. Therefore, 1000 sky maps for every signal strength within a certain range are simulated and the $D^2$-value for each sky map is calculated separately to obtain the $D^2$ distributions. The calculation of the average upper limit at 90\% confidence level assuming zero-signal is used to estimate the sensitivity of the analysis to different astrophysical models apriori. 

As the multipole analysis is applied to a wide range of astrophysical topics, the trial factor of the analysis becomes important. The trial factor raises with each new set of weights used to evaluate the experimental data. For this reason, models with almost similar weights are combined to a common set of weights and only six sets are remaining.

If the signal signatures show up only in the zenith direction of the analysis coordinate system the expansion coefficients $a_l^0$ are more sensitive than the multipole moments $C_l$. The reason is, that the expansion coefficients with $m = 0$ are independent from the azimuth $\phi$ and contain the pure information about the zenith direction $\theta$. A signal only depending on $\theta$ causes only statistical fluctuations but no physical information in the other expansion coefficients. Therefore, the signal has only power in the $a^0_l$. The analysis method stays exactly the same in these cases, except that all $C_l$ are replaced by the $a^0_l$. This is related to the models of neutrinos from the Galactic plane and from sources of the VCV catalog, which show north-south-symmetries of the neutrino signals in Galactic and supergalactic coordinates, respectively. Unlike the multipole moments $C_l$, the $a_l^0$ do not average over different orientations. Therefore, the analysis of the $a_l^0$ strongly depends on the used coordinate system. An example for pulls of $a_l^0$ for the model of a diffuse neutrino flux from the Galactic plane is shown in fig. \ref{sub_fig2}. The characteristic periodic behavior of the pulls is explained by the symmetry properties of the spherical harmonics.

\section{Experimental results}

The experimental data is analyzed in two steps. First, the experimental data is tested for its compatibility with the pure atmospheric neutrino hypothesis. Secondly, the experimental pulls are compared with the expectations for the different investigated neutrino models.

 \begin{figure}[!t]
  \centering
  \includegraphics[width=3.3in]{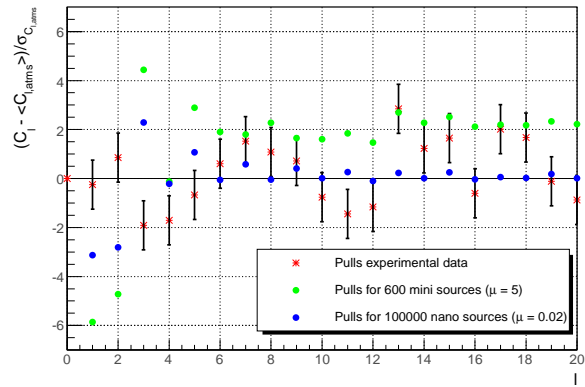}
  \caption{Pull plot for the experimental multipole moments $C_l$. Expected pulls for typical model parameters of isotropic point sources are shown for comparison. The error bars symbolize the statistical fluctuation expected for an atmospheric neutrino sky map.}
  \label{fig3}
 \end{figure}

 \begin{figure}[!t]
  \centering
  \includegraphics[width=3.3in]{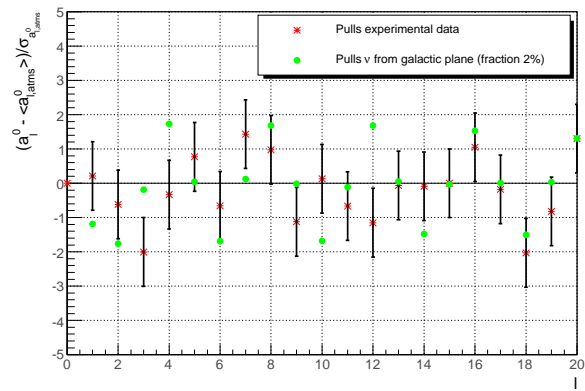}
  \caption{Pull plot for the experimental expansion coefficients $a_l^0$ in Galactic coordinates. Expected pulls for typical parameters of cosmic ray interactions with the Galactic plane are shown for comparison. The error bars symbolize the statistical fluctuation expected for an atmospheric neutrino sky map.}
  \label{fig4}
 \end{figure}

The pulls of the experimental data are shown for the multipole moments $C_l$ in Fig. \ref{fig3} and for the expansion coefficients $a_l^0$ in Galactic coordinates in Fig. \ref{fig4}. To compare the measured data with the expected event distribution, a $D^2$ is calculated for the multipole moments $C_l$ and the expansion coefficients $a_l^0$ for transformations into equatorial, Galactic and supergalactic coordinates separately. As no signal model is tested $\textrm{sgn}_l = w_l = 1$ is assumed. A comparison with the corresponding $D^2$ distributions results in the p-values giving the probability to obtain a $D^2$ which is at least as extreme as the measured one assuming that the pure atmospheric neutrino hypothesis is true (Table \ref{p-values}).

The statistical consistency of $C_l$ and $a_l^0$ in equatorial coordinates with the atmospheric expectation is marginal. Rotating to inclined coordinate systems, e.g. Galactic and supergalactic, the consistency improves. The deviation from the pure atmospheric expectation is not compatible with any of the signal models (see Fig. \ref{fig3}, \ref{fig4} for examples). The discrepancy may be attributed to uncertainties in the theoretical description of the atmospheric neutrino distribution, or to a contribution of unsimulated background of down-going muons mis-reconstructed as up-going, or to the modeling of properties of the AMANDA detector.

  \begin{table}[!h]
  \caption{p-values for the compatibility of experimental data and pure atmospheric neutrino hypothesis.}
  \label{p-values}
  \centering
  \begin{tabular}{|l|c|}
  \hline
   Observable  &  p-value \\
   \hline 
    $C_l$ & 0.02 \\
    $a_l^0$, Equatorial & 0.02 \\
    $a_l^0$, Galactic & 0.15 \\
    $a_l^0$, Supergalactic & 0.70 \\
  \hline
  \end{tabular}
  \end{table}
 
The signal models are tested by calculating the $D^2$-values of the experimental data using the corresponding sign and weight factors. As the observed deviations do not fit any of the investigated signal models the physical model parameters are constrained. Due to the observed systematic effects affecting mainly the multipole moments $C_l$ and the equatorial expansion coefficients $a_l^0$ no limits on the models analyzed in the corresponding coordinate systems (models 1, 2 and 6) are derived. The other models are less affected. The limits given below do not include these systematic effects.

A limit on the source strength assuming the VCV source distribution (model 3) is calculated for those sources closer than 100 Mpc to the Earth. In this model all sources are expected to have the same strength and energy spectrum. For a typical spectral index of $\gamma = 2$ the average source flux is limited by the experimental data to a differential source flux of $\mathrm{d} \Phi / \mathrm{d} E \cdot E^2 \le 1.6 \cdot 10^{-10}\, \mathrm{GeV}\,\mathrm{cm}^{-2}\,\mathrm{s}^{-1}\, \mathrm{sr}^{-1}$ in the energy range between 1.6 TeV and 1.7 PeV.

For the random Galactic sources (model 4), the number of sources is constrained assuming the same source strength and energy spectrum for all sources as well. For a spectral index of $\gamma = 2$, the limit on the number of sources is set by AMANDA to $N_\mathrm{sources} \le 39$ assuming a source strength of $\mathrm{d} \Phi / \mathrm{d} E \cdot E^2 \le 10^{-8}\, \mathrm{GeV}\,\mathrm{cm}^{-2}\,\mathrm{s}^{-1}\, \mathrm{sr}^{-1}$ or $N_\mathrm{sources} \lesssim 4300$ for sources with $\mathrm{d} \Phi / \mathrm{d} E \cdot E^2 \le 10^{-10} \, \mathrm{GeV}\,\mathrm{cm}^{-2}\,\mathrm{s}^{-1}\, \mathrm{sr}^{-1}$. For source fluxes in between the limit can be approximated by assuming linearity between $N_\mathrm{sources}$ and log$\left( \mathrm{d} \Phi / \mathrm{d} E \cdot E^2 \right)$.

\begin{figure}[!t]
  \centering
  \includegraphics[width=3.3in]{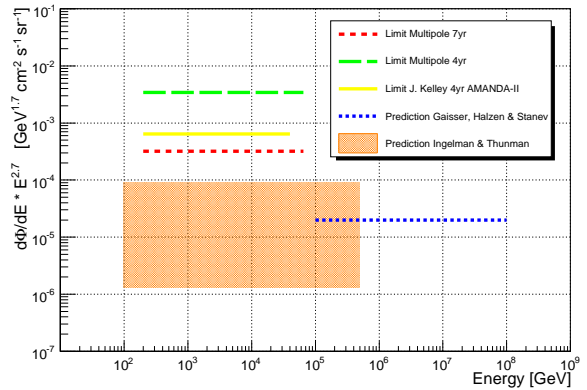}
  \caption{Limit of the 7yr multipole analysis on the diffuse neutrino flux from cosmic ray interactions in the Galactic plane in dependence of the valid energy range. The limit is compared with two other analyses \cite{MUL4YEAR, KELLEY} and two theoretical predictions \cite{GHS, INGTHUN}.}
  \label{fig5}
 \end{figure}

The differential flux limit obtained from the experimental data on the diffuse neutrino flux from cosmic ray interactions in the Galactic plane (model 5) is $\mathrm{d} \Phi / \mathrm{d} E \cdot E^{2.7} \le 3.2 \cdot 10^{-4}\, \mathrm{GeV}^{1.7}\,\mathrm{cm}^{-2}\,\mathrm{s}^{-1}\, \mathrm{sr}^{-1}$. This flux limit is shown in Fig. \ref{fig5} together with the results of two other AMANDA analyses and two theoretical flux predictions. The seven year multipole analysis provides currently the best limit. However, it is still not in reach of the theoretical predictions. \\

\section{Conclusion}

It is shown that the multipole analysis is sensitive to a wide range of physical topics. Its area of application is in particular the field of many weak sources in transition to diffuse fluxes. With the statistics of seven years of AMANDA data and improvements of the analysis technique the method is now restricted by systematic uncertainties in the atmospheric neutrino zenith distribution of the order of a few percent. Transforming to coordinate systems less affected by the equatorial zenith angle such as the Galactic and supergalactic system physical conclusions are still possible. A compatibility of the measurement with the background expectation of atmospheric neutrinos is observed. Current efforts to better understand the observed systematics would allow an application of the multipole analysis on future high statistic IceCube data. \\

\end{document}